\documentclass[
 amsmath,amssymb,
 longbibliography,
 lengthcheck,
]{elsarticle}

\usepackage{graphicx}% Include figure files
\usepackage{dcolumn}% Align table columns on decimal point
\usepackage{epsfig}% Align table columns on decimal point
\usepackage{bm}% bold math
\usepackage{hyperref}% add hypertext capabilities
\usepackage{hhline}% add hypertext capabilities
\begin{document}

\title{Constraints on the $\tilde{H}$ Generalized Parton Distribution 
from Deep Virtual Compton Scattering Measured at HERMES.}
\author{M. Guidal} 
\address{Univ Paris-Sud, Institut de Physique Nucl\'eaire d'Orsay, Orsay,
F-91405}
\date{\today}

\begin{abstract}
We have analyzed the longitudinally polarized proton target asymmetry data of
the Deep Virtual Compton process recently published by the HERMES
collaboration in terms of Generalized Parton Distributions. We have fitted
these new data in a largely model-independent fashion and the procedure 
results in numerical constraints on the $\tilde{H}_\mathrm{Im}$ Compton Form Factor. 
We present its $t-$ and $\xi-$ dependencies. We also find improvement on the 
determination of two other Compton Form Factors, $H_\mathrm{Re}$ and 
$H_\mathrm{Im}$.

%PACS : 13.60.Le, 13.60.Fz, 13.60.Hb
\end{abstract}

%\pacs{13.60.Fz,12.38.Qk}% PACS, the Physics and Astronomy
                             % Classification Scheme.
%\keywords{Suggested keywords}%Use showkeys class option if keyword
                              %display desired

\maketitle

The Deep Virtual Compton Scattering (DVCS) process, i.e.\ the
electroproduction at large virtuality $Q^2$ of a real photon off the
nucleon, is the most favorable channel to access Generalized Parton
Distributions (GPDs). GPDs encode the complex parton (quark and gluon)
substructure of the nucleon, not yet fully calculable from the first
principles of Quantum Chromo-Dynamics (QCD).  GPDs describe, among
many other aspects, the (correlated) spatial and momentum
distributions of the partons in the nucleon (including polarization
degrees of freedom), its quark-antiquark content, they provide a way
to access the orbital momentum contribution of the quarks to the
nucleon's spin, etc.  We refer the reader to
Refs.~\cite{muller,ji,rady,collins,goeke,revdiehl,revrady,barbara,myppnp},
which contain very detailed and quasi-exhaustive reviews on the GPD
formalism and the definitions of some of the variables and notations
that will be employed in the following.

\begin{figure}[htb]
\epsfxsize=9.cm
\epsfysize=10.cm
\epsffile{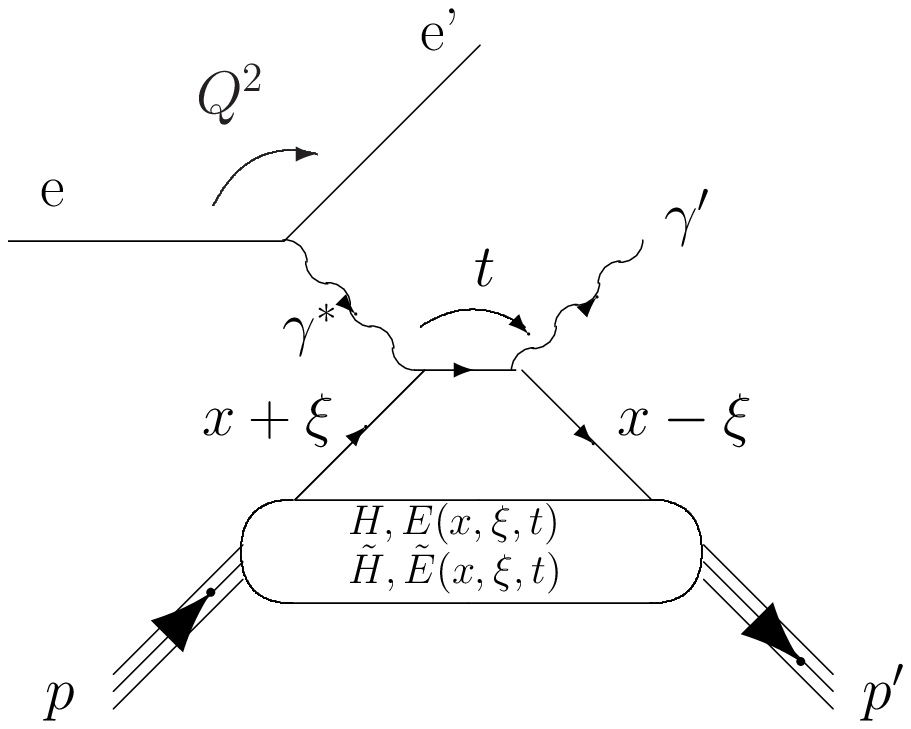}
\vspace{-4.2cm}
\caption{The handbag diagram for the DVCS process off the proton which
  is accessed through the $ep\to e'p'\gamma$ reaction (there is also
  a crossed diagram which is not shown here). In this letter, we focus
  on the proton DVCS process, so that all GPDs and related quantities
  should be understood as proton GPDs in the following.}
\label{fig:dvcs}
\end{figure}

We recall that there are, in the QCD leading twist/leading order 
approximation which is the frame of this study, four independent GPDs
which can be accessed in the DVCS process: $H, E, \tilde{H}$ and $\tilde{E}$. 
They correspond to the various spin and helicity 
orientations of the quark and nucleon in the handbag diagram
of Fig.~\ref{fig:dvcs}. 

These four GPDs depend on three variables $x$, $\xi$ and $t$. The
quantities $x+\xi$ and $x-\xi$ denote the longitudinal momentum
fractions of the initial and final quark (or antiquark) respectively,
in a frame where the nucleon has a large momentum along a certain
direction which defines the longitudinal components.  The variable
$t=(p-p')^2$ is the squared momentum transfer between the initial and
final nucleon with $p$ and $p'$ as four-momenta respectively.
The variable $\xi$ is related to the standard Bjorken variable: 
$\xi\simeq\frac{x_B}{2-x_B}$ in the Bjorken limit as $Q^2\to\infty$,
with $x_B=\frac{Q^2}{2p.q}$ where $q$ is the 
four-momenta of the virtual photon. Only $\xi$ and $t$ can be determined 
experimentally in DVCS. Thus, only Compton Form Factors (CFFs), which are weighted 
integrals of GPDs over $x$ or combinations of GPDs at the line
$x=\xi$ and which therefore depend only on the two kinematic variables
$\xi$ and $t$, can actually be measured in DVCS experiments. 
Eight CFFs arise from the decomposition of the DVCS amplitude into real and
imaginary parts and, following our notations and
definitions introduced in Refs.~\cite{fitmick,fithermes,fittsa}, we
call them: $H_\mathrm{Re}, E_\mathrm{Re}, \tilde{H}_\mathrm{Re},
\tilde{E}_\mathrm{Re}, H_\mathrm{Im}, E_\mathrm{Im},
\tilde{H}_\mathrm{Im}$ and $\tilde{E}_\mathrm{Im}$. The CFFs
with the ``$\mathrm{Re}$" index refer to the weighted integrals of GPDs
over $x$ and those with the ``$\mathrm{Im}$" index refer to the GPDs at the 
particular point $x=\xi$. We stress that what we call Compton Form Factor
does not correspond exactly to the original definition of Ref~\cite{kirch}.
In this latter article, CFFs are, up to minus signs and $\pi$ factors, the 
complex sum of our ``$\mathrm{Re}$" and ``$\mathrm{Im}$" CFFs.

In Refs.~\cite{fitmick,fithermes,fittsa}, we have developed a largely
model-independent fitting procedure which, at a given experimental
($\xi$, $-t$) kinematic point, takes the CFFs as free parameters and
extracts them from DVCS experimental observables using the well
established DVCS theoretical amplitude~\cite{kirch,vgg1}. This task
is not trivial. Firstly, one has to fit eight parameters from a
limited set of data and observables, which leads in general to an
under-constrained problem.  However, as some observables are in
general dominated by a few particular CFFs, one can manage, in some
cases, to extract some specific CFFs.  Secondly, there is, in addition
to the particular DVCS process of
direct interest, another mechanism which contributes to the $ep\to
e'p'\gamma$ process. This is the Bethe-Heitler (BH) process where the
final state photon is radiated by the incoming or scattered electron
and not by the nucleon itself. This latter reaction carries no
useful information about GPDs. The BH process interferes with
DVCS and in some parts of the phase space has a cross section which
dominates the DVCS one. It can therefore mask or ``distort" (favorably or unfavorably)
the DVCS and GPD signals and it is crucial to properly take it into
account. The BH process is however relatively precisely known and
calculable given the nucleon form factors.

With our fitting algorithm, we have managed to determine in previous 
works~\cite{fitmick,fithermes,fittsa}, within average uncertainties of the 
order of 30\%~:
\begin{itemize}
\item the $H_\mathrm{Im}$ and $H_\mathrm{Re}$ CFFs. at
  $<x_B>\approx0.36$, and for several $t$ values, by
  fitting~\cite{fitmick} the JLab Hall A proton DVCS beam-polarized
  and unpolarized cross sections~\cite{franck},
\item the $H_\mathrm{Im}$ and $\tilde{H}_\mathrm{Im}$ CFFs, at
  $<x_B>\approx 0.35$ and $<x_B>\approx 0.25$, and for several $t$
  values, by fitting~\cite{fittsa} the JLab CLAS proton DVCS
  beam-polarized and longitudinally polarized target spin
  asymmetries~\cite{fx,chen},
\item the $H_\mathrm{Im}$ and $H_\mathrm{Re}$ CFFs, at $<x_B>\approx
  0.09$, and for several $t$ values, by fitting~\cite{fithermes} a
  series (seventeen) of HERMES beam-charge, beam-polarized and transversely
  polarized target spin asymmetry moments~\cite{ave,hermes}.
\end{itemize}

Now, the HERMES collaboration has recently published~\cite{hermesaul}
two new proton DVCS observables: the single spin asymmetry with a
longitudinally polarized proton target and the double spin asymmetry with a
polarized positron beam and a longitudinally polarized proton target (the 
direction of the virtual photon defines the longitudinal axis here). These 
two independent observables are presented in the form of moments in
Ref.~\cite{hermesaul,davidthesis}, denoted as~: $A_\mathrm{UL}^{\sin\phi}$,
$A_\mathrm{UL}^{\sin2\phi}$, $A_\mathrm{UL}^{\sin3\phi}$,
$A_\mathrm{LL}^0$, $A_\mathrm{LL}^{\cos\phi}$ and
$A_\mathrm{LL}^{\cos2\phi}$. In this notation, the first index of the
asymmetry $A$ refers to the polarization of the beam (``U" for
unpolarized and ``L" for longitudinally polarized) and the second one
to the polarization of the target (``L" for a longitudinally polarized
target).  The superscript refers to the harmonic dependence of the
asymmetries. $\phi$ is the azimuthal angle between the leptonic 
and hadronic planes~\cite{hermesaul,davidthesis}.

In this letter, we study what new information these additional
observables can bring.  It is well known~\cite{kirch,fitmick,fittsa}
that the longitudinally polarized target spin observable is
predominantly sensitive to $\tilde{H}_\mathrm{Im}$. We thus expect to
extract numerical constraints on this particular CFF for the first
time at HERMES kinematics. Our procedure consists in fitting these six
new moments \underline{in addition} to the HERMES other (seventeen) 
moments previously mentioned, related to the beam-charge,
beam-polarized and transversely polarized target spin asymmetries
(many of these moments being zero in the leading twist DVCS
approximation). We recall that we already fitted in
Ref.~\cite{fithermes} these seventeen moments but no
convergence of the $\tilde{H}_\mathrm{Im}$ CFF towards some
well-defined domain could be observed (in contrast with the
${H}_\mathrm{Im}$ and ${H}_\mathrm{Re}$ CFFs).

The parameters to be fitted are the CFFs and the function to be minimized is:

\begin{equation}
\chi^2=\sum_{i=1}^{n}
\frac{(A^\mathrm{theo}_i-A^\mathrm{exp}_i)^2}{(\delta\sigma^\mathrm{exp}_i)^2}
\label{eq:chi2}
\end{equation}

\noindent where $i$ runs over all the twenty-three (seventeen + six)
HERMES asymmetry moments previously mentioned, $A^\mathrm{theo}$ is
the theoretical asymmetry moment calculated from the sum of the
leading twist/leading order DVCS amplitude and of the exact BH
amplitude, $A^\mathrm{exp}$ is the corresponding HERMES experimental
value and $\delta\sigma_\mathrm{exp}$ is its associated experimental
error bar. We have used MINUIT and MINOS~\cite{james} to 
carry out the $\chi^2$ minimization and to determine the uncertainties 
on the fitted parameters.

We mentioned earlier that there are in principle eight CFFs appearing
in the DVCS process. As in Refs.~\cite{fitmick,fithermes,fittsa}, we
have actually considered only seven CFFs as we have set
$\tilde{E}_\mathrm{Im}$ to zero, guided by theoretical
considerations. The $\tilde{E}$ GPD is indeed in general associated to
the pion pole exchange in the $t$-channel whose amplitude is real. 
We stress that this is essentially the only model assumption in our procedure. 

Also, following what we have done and explained in details in 
Refs.~\cite{fitmick,fithermes,fittsa}, another feature entering our fitting 
procedure is that we contrain the domain of the fitting parameters 
(i.e.\ the CFFs) to be $\pm$5 times a set of ``reference" CFFs. Without any bounding, 
our fits which are in general underconstrained, would not converge. These 
reference CFFs are the ``VGG" CFFs. VGG~\cite{vgg1,goeke,gprv} is a well-known 
and widely used model which provides an acceptable first approximation of the 
CFFs, as shown in our previous studies~\cite{fitmick,fithermes,fittsa} and as 
will be confirmed furtherdown in the present work. We recall that some GPDs
have to satisfy a certain number of normalization constraints. These
are all fulfilled by the VGG model.  It should be clear that $\pm$5
times the VGG CFFs make up extremely conservative bounds and that this
bounding can barely be considered as model-dependent. 

\begin{figure}[htb]
\epsfxsize=13.cm
\epsfysize=9.cm
\epsffile{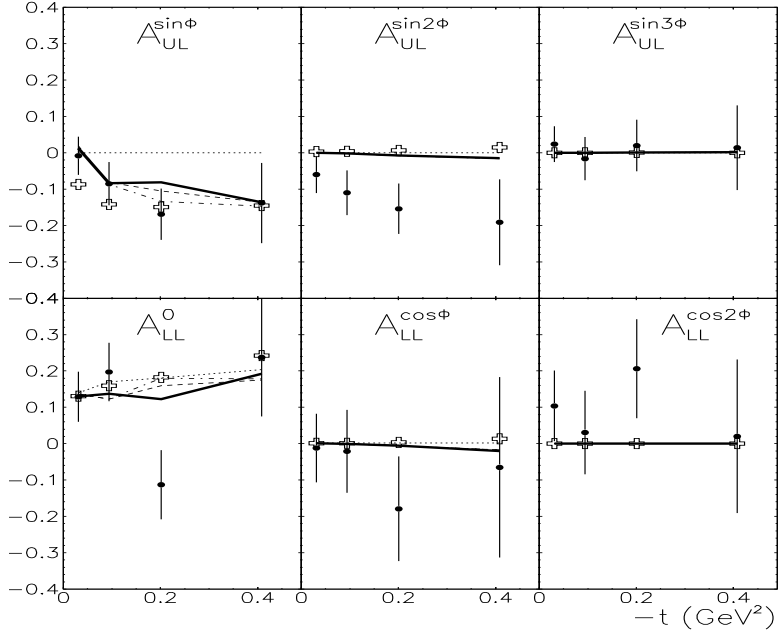}
%\vspace{-4.2cm}
\caption{The six HERMES $A_\mathrm{UL}$ and $A_\mathrm{LL}$ moments,
  as a function of $-t$, which are fitted simultaneously (along with
  seventeen other moments previously published by HERMES). The solid
  circles are the data points of Ref.~\cite{hermesaul}.  The thick
  solid line is the result of our fit which includes all the other
  seventeen moments of Refs.~\cite{ave,hermes}. The dashed
  line shows the result of our fit \underline{excluding} the three
  $A_\mathrm{LL}$ moments (i.e.\ only twenty moments are fitted
  simultaneously). The dot-dashed line shows the result of our fit
  \underline{excluding} also the $A_\mathrm{\{Ux,I\}}^{\sin\phi}$
  moment of Ref.~\cite{ave} (i.e.\ nineteen moments are fitted
  simultaneously).  The results of our fits are calculated at the
  experimental bin centres and connected by straight lines for
  visibility only. The dotted line is the result of the BH alone
  calculation. The empty crosses are the predictions of the VGG
  model.}
\label{fig:hermes2}
\end{figure}

Under these largely model-independent conditions, we then obtain the
fits, as a function of $t$, of the six HERMES $A_\mathrm{UL}$ and $A_\mathrm{LL}$ 
moments shown in Fig.~\ref{fig:hermes2}.  The thick solid line is the result
of the fit of the twenty-three HERMES moments: the seventeen from
Refs.~\cite{ave,hermes} (not shown here) and the six from
Ref.~\cite{hermesaul} shown in this figure.  Within our leading twist/leading 
order DVCS framework, only three of the moments in
Fig.~\ref{fig:hermes2}, i.e.\ $A_\mathrm{UL}^{\sin\phi}$, $A_\mathrm{LL}^0$ 
and $A_\mathrm{LL}^{\cos\phi}$ can be significantly different from zero.
The other three moments $A_\mathrm{UL}^{\sin2\phi}$, 
$A_\mathrm{UL}^{\sin3\phi}$ and $A_\mathrm{LL}^{\cos2\phi}$ are higher
twist contributions. The data for the latter two moments are, in
general, compatible with zero and therefore quite well fitted by our
code. However, the data for the $A_\mathrm{UL}^{\sin2\phi}$ moment
are systematically significantly different from zero.  This is
impossible to achieve within our framework. Out of all our previous
studies~\cite{fitmick,fithermes,fittsa}, this is the first observable
that we systematically cannot well reproduce with our fitting code. In
other words, this is the first significant and systematic DVCS higher
twist sign that we encouter. This was of course also noted in the
HERMES publication~\cite{hermesaul}. However, let us mention that,
while in Fig.~\ref{fig:hermes2}, we show the $t$-dependence of the
$A_\mathrm{UL}$ and $A_\mathrm{LL}$ moments at (almost) fixed $<x_B>$, in
Ref.~\cite{hermesaul}, the $x_B$ dependence at (almost) fixed $<-t>$ is also
shown. There, it can then be observed that only one
$A_\mathrm{UL}^{\sin2\phi}$ data point, among four, at an intermediate
$<x_B>$ value ($<x_B>$=0.084), is significantly different from
zero. This rather ``local" deviation is intriguing and could hint that a
statistical fluctuation might not be completely out of the
question. Otherwise, one has to conceive a non-obvious mechanism which provides
such a sharp and local rise of $A_\mathrm{UL}^{\sin2\phi}$ at that
particular $<x_B>$ value.

This poor reconstruction of the $A_\mathrm{UL}^{\sin2\phi}$ moment being
noticed and unexplained, we now focus on the three
$A_\mathrm{UL}^{\sin\phi}$, $A_\mathrm{LL}^0$ and $A_\mathrm{LL}^{\cos\phi}$
moments which are ``allowed" to be significantly different from zero at 
leading twist DVCS. We first note in Fig.~\ref{fig:hermes2} that
$A_\mathrm{LL}^{\cos\phi}$ is very small although it is leading twist.
It is indeed largely domainted by BH which gives this moment essentially
equal to zero. 
Now, turning to the two other leading twist moments, $A_\mathrm{UL}^{\sin\phi}$ 
and $A_\mathrm{LL}^0$, which experimentally do turn out to be significantly 
different from zero, we see that they are well fitted for three out of 
four $t$-values. Indeed, our fit (thick 
solid line) misses the third $t$ point, i.e.\ at $<-t>=0.201$ GeV$^2$, for 
both observables. Once again, a strong ``local" discontinuity occurs for
$A_\mathrm{LL}^0$, being negative for this particular value $t$ value
while being positive for the other three $t$ values. This might point
to another local and singular statistical effect.  The dashed line in
Fig.~\ref{fig:hermes2} shows the result of our fit if we remove the
$A_\mathrm{LL}$ moments. We then note that the fit of
$A_\mathrm{UL}^{\sin\phi}$ is improved.

Still, the dashed line is not passing exactly through the central
value of $A_\mathrm{UL}^{\sin\phi}$ at $<-t>=0.201$ GeV$^2$. We recall
that we fit simultaneously twenty-three moments and not only the
$A_\mathrm{UL}$ and $A_\mathrm{LL}$ moments. This means that another
observable must outweigh $A_\mathrm{UL}^{\sin\phi}$ and push the fit
away from the $A_\mathrm{UL}^{\sin\phi}$ data point at that particular 
$t$ value. We will identify this other observable furtherdown and will 
then describe the dot-dashed curve which does correctly fit all
$A_\mathrm{UL}^{\sin\phi}$ data points.

In Fig.~\ref{fig:hermes2}, the dotted curve shows the result of the
calculation with only the BH contribution. It is essentially zero for all
$A_\mathrm{UL}$ and $A_\mathrm{LL}$ moments but $A_\mathrm{LL}^0$. 
The dotted curve
shows that the main contribution to this latter moment comes basically
from BH alone and that the DVCS contribution to this observable is
very small.  The empty crosses in Fig.~\ref{fig:hermes2} show the VGG
prediction which, except for the puzzling higher twist
$A_\mathrm{UL}^{\sin2\phi}$, gives a relatively good overall
description of the data.

\begin{figure}
\epsfxsize=13.cm
\epsfysize=12.cm
\epsffile{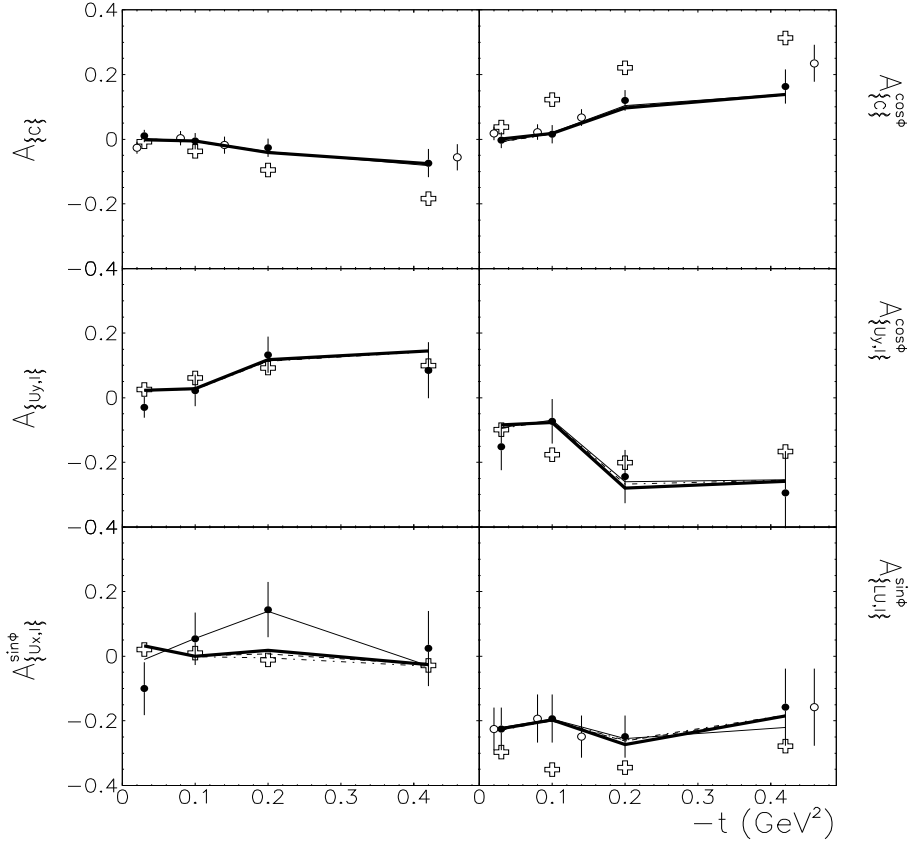}
%\vspace{-4.2cm}
\caption{Six out of seventeen moments determined at HERMES, other
  than those of Fig.~\ref{fig:hermes2}, which are fitted simultaneously
  in this work as a function of $-t$.  For the two $A_\mathrm{\{C\}}$
  asymmetry moments, the solid circles show the HERMES data of
  Refs.~\cite{ave,hermes} and the open circles show the HERMES
  data of Ref.~\cite{hermes}. For the three
  $A_\mathrm{\{U(x,y)\}}$ asymmetry moments, the solid circles show
  the HERMES data of Refs.~\cite{ave,hermes}. For the
  $A_\mathrm{\{LU\}}$ asymmetry moment, the open circles show the
  HERMES data of Ref.~\cite{hermes} and the solid circles show these
  SAME data offset to the kinematics of Ref.~\cite{ave} (and of
  Ref.~\cite{hermesaul}), so as to fit all (twenty-three) moments
  simultaneously at the same kinematics.  In other words, the solid
  circles in all panels show the data point which have actually been
  fitted.  The thick solid line is the result of our fit including all
  twenty-three moments.  The thin solid line is the result of our fit,
  previously published in Ref.~\cite{fithermes},
  i.e.\ \underline{excluding} the three $A_\mathrm{UL}$ and the three
  $A_\mathrm{LL}$ moments (of Fig.~\ref{fig:hermes2}), i.e.\ seventeen
  moments have been fitted.  In most cases, the thin solid line overlaps
  with the thick solid line and cannot be distinguished.
  The dashed line (also barely visible) shows
  the result of our fit \underline{excluding} the three
  $A_\mathrm{LL}$ moments (i.e.\ twenty moments are fitted). The
  dot-dashed line (mostly visible on the $A_\mathrm{\{Ux,I\}}^{\sin\phi}$
  panel) shows the result of our fit \underline{excluding}
  in addition the $A_\mathrm{\{Ux,I\}}^{\sin\phi}$ moment of Ref.~\cite{ave}
  (i.e.\ nineteen moments are fitted).  The results of our fits are
  calculated at the experimental bin centres and connected by straight
  lines for visibility only. The empty crosses are the predictions of
  the VGG model.}
\label{fig:hermes3}
\end{figure}

We now show in Fig.~\ref{fig:hermes3} six of the other seventeen
observables that we simultaneously fit with the $A_\mathrm{UL}$
and $A_\mathrm{LL}$ moments of Fig.~\ref{fig:hermes2}. We recall that
this set of seventeen moments was studied in detail in
Ref.~\cite{fithermes}.  The six moments diplayed in
Fig.~\ref{fig:hermes3}, i.e.\ $A_\mathrm{\{C\}},
A_\mathrm{\{C\}}^{\cos\phi}, A_\mathrm{\{Uy,I\}},
A_\mathrm{\{Uy,I\}}^{\cos\phi}, A_\mathrm{\{Ux,I\}}^{\sin\phi}$ and
$A_\mathrm{\{LU,I\}}^{\sin\phi}$ are the moments which can be
significantly different from zero at leading twist DVCS, out of the
seventeen published by HERMES in Refs.~\cite{ave,hermes}. 
These observables actually originate from two independent analyses
(although they bear on the same data set).
Ref.~\cite{ave} has extracted the $A_\mathrm{\{C\}}$ and
$A_\mathrm{\{UT\}}$ asymmetries while Ref.~\cite{hermes} have extracted
the $A_\mathrm{\{LU\}}$ asymmetries (as well as, simultaneously, the
$A_\mathrm{\{C\}}$ asymmetries also). The two analyses have different
binnings and thus slightly different central $-t$ (and $x_B$)
values. As was explained in details in Ref.~\cite{fithermes}, in order
to be able to fit simultaneously all observables, we have considered
that all data were taken at the four $-t$ points of
Ref.~\cite{ave}. These four $-t$ points are actually the same as in
Ref.~\cite{hermesaul} (and thus as in Fig.~\ref{fig:hermes2}). The actual
fitted data are therefore the solid circles in Fig.~\ref{fig:hermes3}. 
A slight uncertainty for the
$A_\mathrm{\{LU\}}$ asymmetries is thus introduced since they are
therefore not calculated at the exact kinematics at which they were
measured. Given the uncertainties on our final results, which we will
present shortly, we consider this effect as negligible.

Similar to Fig.~\ref{fig:hermes2}, the thick solid line in
Fig.~\ref{fig:hermes3} shows the result of our fit when all
twenty-three HERMES moments are included in the fit,
i.e.\ \underline{with} the new $A_\mathrm{UL}$ and $A_\mathrm{LL}$
moments. The thin solid line in Fig.~\ref{fig:hermes3} shows the
results that we previously published in Ref.~\cite{fithermes},
i.e.\ \underline{without} the new $A_\mathrm{UL}$ and $A_\mathrm{LL}$
moments. Except for $A_\mathrm{\{Ux,I\}}^{\sin\phi}$, the thick and
solid lines are essentially superimposed, which shows that the
introduction of the $A_\mathrm{UL}$ and $A_\mathrm{LL}$ moments in our
fit did not in general strongly affect our previous results, as expected. 
However, there is a striking difference for
$A_\mathrm{\{Ux,I\}}^{\sin\phi}$ at the third $t$ point, i.e.\ at
$<-t>=0.201$ GeV$^2$. This is precisely the $-t$ value for which we
previously observed some problem for $A_\mathrm{UL}^{\sin\phi}$ and
$A_\mathrm{LL}^0$ (see Fig.~\ref{fig:hermes2}). Within our fitting
algorithm (we recall, based on the leading twist DVCS assumption), it
thus doesn't appear possible to fit \underline{simultaneously} these
three asymmetry moments,
i.e.\ $A_\mathrm{\{Ux,I\}}^{\sin\phi}$, $A_\mathrm{UL}^{\sin\phi}$ and
$A_\mathrm{LL}^0$.  Indeed, the \underline{thin} solid line in
Fig.~\ref{fig:hermes3} (i.e.\ the fit \underline{without} the
$A_\mathrm{UL}$ and $A_\mathrm{LL}$ moments) perfectly fits
$A_\mathrm{\{Ux,I\}}^{\sin\phi}$ while the \underline{thick} solid
line in Figs.~\ref{fig:hermes2} and~\ref{fig:hermes3} (i.e.\ the fit
\underline{with} the $A_\mathrm{UL}$ and $A_\mathrm{LL}$ moments
included) misses both $A_\mathrm{\{Ux,I\}}^{\sin\phi}$ and
$A_\mathrm{UL}^{\sin\phi}$. In other words, including
$A_\mathrm{UL}^{\sin\phi}$ in the data to be fitted spoils the fit of
$A_\mathrm{\{Ux,I\}}^{\sin\phi}$ (at $<-t>=0.201$ GeV$^2$). The
uncertainties of $A_\mathrm{\{Ux,I\}}^{\sin\phi}$ and
$A_\mathrm{UL}^{\sin\phi}$ at $<-t>=0.201$ GeV$^2$ are respectively
$\approx$ 60\% and $\approx$ 40\%. Thus, the minimization procedure
finds some sort of ``intermediate" solution in order to accomodate
both data points when both moments, $A_\mathrm{\{Ux,I\}}^{\sin\phi}$
and $A_\mathrm{UL}^{\sin\phi}$, are included in the fit (thick solid
line in Figs.~\ref{fig:hermes2} and~\ref{fig:hermes3}).

In order to better understand this issue, we have removed the
$A_\mathrm{\{Ux,I\}}^{\sin\phi}$ and $A_\mathrm{LL}$ moments of
our fit which seem to pose problems.
The result of this fit is the dot-dashed curve in Figs.~\ref{fig:hermes2} 
and~\ref{fig:hermes3}. $A_\mathrm{UL}^{\sin\phi}$ is now very well fitted,
in particular the third $t$ point at $<-t>=0.201$ GeV$^2$.
Of course, since $A_\mathrm{\{Ux,I\}}^{\sin\phi}$ was not included in the fit,
it is not particularly well fitted, in particular the problematic
third $t$ point at $<-t>=0.201$ GeV$^2$ in Fig.~\ref{fig:hermes3}.
To summarize this discussion, the inclusion of 
$A_\mathrm{UL}^{\sin\phi}$ in the fit seems to spoil the fit of 
$A_\mathrm{\{Ux,I\}}^{\sin\phi}$ and, oppositely, the inclusion of
$A_\mathrm{\{Ux,I\}}^{\sin\phi}$ in the fit seems to spoil the fit 
of $A_\mathrm{UL}^{\sin\phi}$.

As we will see shortly, the precise value of
$A_\mathrm{UL}^{\sin\phi}$ is going to directly impact the value of
the $\tilde{H}_\mathrm{Im}$ CFF. It would therefore be important to
clarify which one of the two fitting curves, i.e.\ the thick solid one
or the dot-dashed one in Figs.~\ref{fig:hermes2}
and~\ref{fig:hermes3}, one should actually consider.  We cannot
decide alone which data point, between
$A_\mathrm{\{Ux,I\}}^{\sin\phi}$ and $A_\mathrm{UL}^{\sin\phi}$ (at
$<-t>=0.201$ GeV$^2$), is the ``most correct". However, we can notice
that the VGG prediction (the empty crosses in Figs.~\ref{fig:hermes2}
and~\ref{fig:hermes3}) which, in general, gives a decent description
of the data, seems to favor $A_\mathrm{\{Ux,I\}}^{\sin\phi}$ values
close to zero, which differs significantly with the experimental data point at
$<-t>=0.201$ GeV$^2$.

Now that we have compared our fit curves to the data, let us examine
the values of the fitted CFFs which come out of the minimization
procedure. Three CFFs $H_\mathrm{Re}$, $H_\mathrm{Im}$ and
$\tilde{H}_\mathrm{Im}$ come out of our fitting procedure with finite
MINOS uncertainties and stable central values. The other four fitted CFFs did
not converge to some well defined value or domain: either their
central value reached the boundaries of the allowed domain of
variation ($\pm 5$ times the VGG value) or MINOS could not reach the
$\chi^2$+1 value and thus we could not well define the associated uncertainty. The
explanation for which some particular CFFs do converge and do come out
of the fits within well defined and delimited domains is that some
observables are, for dynamical or kinematical reasons, particularly
sensitive to some specific CFFs. For instance, it is well
established~\cite{fitmick,kirch,davidthesis} that DVCS charge asymmetries are in
general mostly sensitive to $H_\mathrm{Re}$, beam single spin asymmetries to
$H_\mathrm{Im}$ and, particularly related to the present work,
longitudinally polarized target single spin asymmetries to
$\tilde{H}_\mathrm{Im}$.

The central values for the $H_\mathrm{Re}$ and $H_\mathrm{Im}$ CFFs
that we obtain in this work, are almost unchanged compared to the ones we 
obtained in Ref.~\cite{fithermes}, where the six $A_\mathrm{UL}$ and 
$A_\mathrm{LL}$ moments were not included. The important difference is that 
now the $\tilde{H}_\mathrm{Im}$ CFF does converge to a well-defined
value. This could of course be anticipated given the previously
mentioned sensitivity of the $A_\mathrm{UL}$ moments to this
particular CFF. We first discuss the $\tilde{H}_\mathrm{Im}$ CFF and
will come back a few paragraphs below to the $H_\mathrm{Re}$ and
$H_\mathrm{Im}$ CFFs and see the improvement gained in their
determination due to the introduction of the $A_\mathrm{UL}$ and
$A_\mathrm{LL}$ moments in the fit. 

\begin{figure}[htb]
\epsfxsize=13.cm
\epsfysize=12.cm
\epsffile{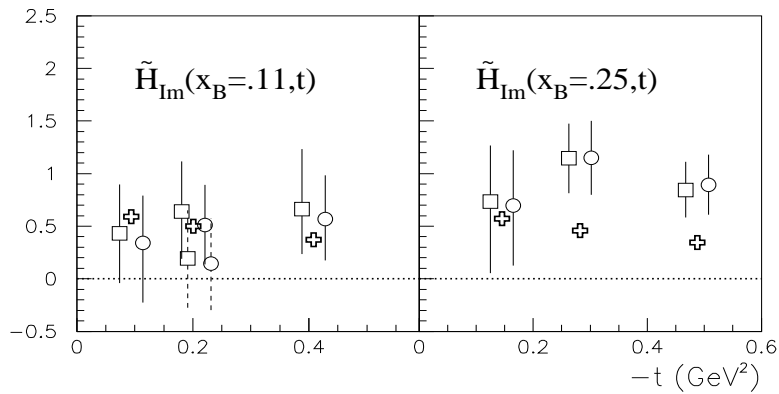}
\vspace{-5.5cm}
\caption{The $t$-dependence of the $\tilde{H}_\mathrm{Im}$ CFF,
  extracted from our fits (left: at HERMES kinematics; right: at CLAS
  kinematics). The empty squares (circles) show our results when the
  boundary values of the domain over which the CFFs are allowed to
  vary is 5 (3) times the VGG reference values. In the left panel, at
  $<-t>=0.201$ GeV$^2$, the set of points with the dashed error bars
  are the results of our fit when all twenty-three moments are
  included. The set of points with the solid error bars are the
  results of our fit when $A_\mathrm{\{Ux,I\}}^{\sin\phi}$ and the
  $A_\mathrm{LL}$ moments are
  \underline{excluded} from the fit (i.e.\ only nineteen moments are
  fitted). The empty crosses indicate the VGG prediction.}
\label{fig:hermes1}
\end{figure}

We show in the left panel of Fig.~\ref{fig:hermes1} the resulting
values of $\tilde{H}_\mathrm{Im}$ that we obtain and
which are therefore an original result. We made the fits for the four
$t$ values of Figs.~\ref{fig:hermes2} and~\ref{fig:hermes3}.  However,
we display in Fig.~\ref{fig:hermes1} $\tilde{H}_\mathrm{Im}$ for only
the three largest $<-t>$ values, i.e.\ $<-t>$=0.094, 0.201 and 0.408
GeV$^2$. Indeed, the MINOS uncertainties on $\tilde{H}_\mathrm{Im}$ were at
the level of 100\% for the smallest $<-t>$ point. We note that at this
(very small) $t$ value, $A_\mathrm{UL}^{\sin\phi}$=-.008 $\pm$ 0.051
$\pm$ 0.012, i.e.\ it is close to zero with, consequently, 
a very important uncertainty.  
In Fig.~\ref{fig:hermes1}, following our convention used in
Ref.~\cite{fittsa}, the empty squares show our results for
$\tilde{H}_\mathrm{Im}$ when the CFFs are limited to vary within
$\pm$5 times the VGG reference values while the open circles show
these results for boundary values equal to $\pm$3 times these same VGG
reference values. The empty square and circle symbols have been
slightly offset from the central $t$ values, left and right
respectively, for sake of visibility.  The uncertainties that we
obtain on our fitted CFFs have in general two origins.  One of course
is related to the statistical precision of the data that are fitted.
The other one stems from the correlation between the fitted
parameters.  This latter cause of uncertainty reflects the potential 
influence of all the other
CFFs. We recall that, in order to be as model-independent as possible,
the essence of our approach is essentially (i.e.\ except for
$\tilde{E}_\mathrm{Im}$) to make no assumption on the value of any of
the CFFs. Then, of course, the smaller the domain of variation allowed
for the CFFs (i.e.\ $\pm$3 times compared to $\pm$5 times the VGG
reference values), the smaller the error bars on the ``convergent"
CFFs due to this effect.  This is what we observed in our previous
studies~\cite{fitmick,fithermes,fittsa}.  We can note in
Fig.~\ref{fig:hermes1} that there is not a strong difference between
the values of the relative error bars of the two cases considered here,
i.e.\ $\pm$3 times and $\pm$5 times the VGG reference values. This is
a sign that these error bars have mostly a statistical origin (we note
that three out of the four $A_\mathrm{UL}^{\sin\phi}$ that we fit have
an experimental uncertainty of more than 80\%, see Fig.~\ref{fig:hermes2}).
This difference will be more pronounced when we will look at the
$H_\mathrm{Re}$ CFF furtherdown.

At $<-t>=0.201$ GeV$^2$, we display two sets of values in the
left panel of Fig.~\ref{fig:hermes1}. The values
with the dashed error bars correspond to the fit when
$A_\mathrm{\{Ux,I\}}^{\sin\phi}$ (and the six $A_\mathrm{UL}$ and
$A_\mathrm{LL}$ moments) is included. We saw in
Figs.~\ref{fig:hermes2} and~\ref{fig:hermes3} (thick solid line) that
then $A_\mathrm{UL}^{\sin\phi}$ is underestimated at that particular
$t$ value.  As a consequence, it can be deduced that in this case
$\tilde{H}_\mathrm{Im}$ will also be underestimated. Hence the value
of $\tilde{H}_\mathrm{Im}$ gets close to zero in
Fig.~\ref{fig:hermes1} at $<-t>=0.201$ GeV$^2$ (empty square and
circle with dashed error bars). Now, if one excludes
$A_\mathrm{\{Ux,I\}}^{\sin\phi}$ (and the three $A_\mathrm{LL}$ moments)
from the fit, this yields, as we saw,
the dot-dashed curves in Figs.~\ref{fig:hermes2}
and~\ref{fig:hermes3}. $A_\mathrm{UL}^{\sin\phi}$ is then correctly
fitted and, as a consequence, $\tilde{H}_\mathrm{Im}$ becomes
larger. In Fig.~\ref{fig:hermes1}, this case corresponds to the empty
square and circle points with the solid line error bars.  
We have a couple of (disputable) arguments which tend to make us think that
this latter fit is more trustworthy: firstly, the relative 
uncertainty on $A_\mathrm{UL}^{\sin\phi}$ is a bit less than on 
$A_\mathrm{\{Ux,I\}}^{\sin\phi}$ tending to give more credit
to the former moment and, secondly, as we mentionned earlier, the former moment 
is more consistent with the VGG predictions. Additionally, the sort of structure
in $A_\mathrm{\{Ux,I\}}^{\sin\phi}$ with some local trend to rise at $<-t>=0.201$ 
GeV$^2$ is not obvious to explain in a GPD model. We nevertheless show both results 
in Fig.~\ref{fig:hermes1} with, thus, a leaning for the solid line error bar
points. Except for
the kinematic point at $<-t>=0.201$ GeV$^2$ that we just discussed,
the central values of the fitted $\tilde{H}_\mathrm{Im}$ CFFs are in
very good agreement with or without the
$A_\mathrm{\{Ux,I\}}^{\sin\phi}$ and $A_\mathrm{LL}$ moments included in the fit.
We therefore don't show the dashed error bars points for the other $t$ values.

In the right panel of Fig.~\ref{fig:hermes1}, we have displayed the
values of $\tilde{H}_\mathrm{Im}$ that we extracted in a previous
work~\cite{fittsa} from the simultaneous fit of the DVCS
longitudinally polarized target single spin asymmetries and the beam single spin
asymmetries measured by the CLAS collaboration~\cite{fx,chen},
i.e.\ at larger $x_B$. We thus obtain, for the first time, an $x_B$-
(or $\xi$-) dependence of $\tilde{H}_\mathrm{Im}$.  Given the
relatively large error bars that we have obtained, it is difficult to
draw clearcut conclusions. Nevertheless,
considering for the HERMES case only the points with the solid error
bars in Fig.~\ref{fig:hermes1}, we observe some similar trends for the
$t$-dependence between the HERMES and CLAS kinematics. If one focuses only 
on the central values, $\tilde{H}_\mathrm{Im}$ tends to go to zero as 
$-t$ goes to zero. Then, some maximum seems to show for $-t$ between 
0.2 and 0.3 GeV$^2$ before a trend to decrease again as $\mid -t\mid$ increases. 
Considering the rather large uncertainties, a flat $t$-dependence 
cannot be excluded either. In any case, there doesn't seem to be 
any strong $t$-dependence for $\tilde{H}_\mathrm{Im}$
(in contrast to standard proton -electromagnetic- form factors).

The $x_B$ dependence doesn't appear to be very strong either. The central values 
of $\tilde{H}_\mathrm{Im}$ tend to show a slow decrease between CLAS and 
HERMES kinematics. This is corroborated by the VGG predictions (empty crosses
in Fig.~\ref{fig:hermes1}) which show little variation of
$\tilde{H}_\mathrm{Im}$ between the two $x_B$ values. This is in
contrast to ${H}_\mathrm{Im}$ which, at fixed $t$, tends to show some
rising behavior as $x_B$ decreases~\cite{fittsa}. In the
comparison with the VGG model, regarding the $t-$dependence, we
however note that there is no decrease of $\tilde{H}_\mathrm{Im}$ as
$-t$ goes to zero. This difference of behavior as $-t$ goes to zero
between the fitted $\tilde{H}_\mathrm{Im}$ and the VGG prediction can
actually be inferred from Fig.~\ref{fig:hermes2} where it can be seen,
in the $A_\mathrm{UL}^{\sin\phi}$ panel, that the empty crosses
overestimate (in absolute value) the data at small $\mid -t\mid$. This explains
that the VGG $\tilde{H}_\mathrm{Im}$ is also overestimated at small
$\mid -t\mid$ compared to the fitted $\tilde{H}_\mathrm{Im}$. As a consequence,
the VGG $\tilde{H}_\mathrm{Im}$ does not show the fall-off trend as
$-t$ goes to zero.

The $t$-dependence of GPDs can be interpreted as a reflection of the
spatial distribution of some
charge~\cite{Burkardt:2000,Diehl:2002he,pire}.  Physically, the smoother
$t$-dependence of $\tilde{H}_\mathrm{Im}$ compared to
${H}_\mathrm{Im}$ could then suggest that the axial charge has
a more narrow distribution than the nucleon than the electromagnetic 
charge.

\begin{figure}[htb]
\epsfxsize=11.cm
\epsfysize=12.cm
\epsffile{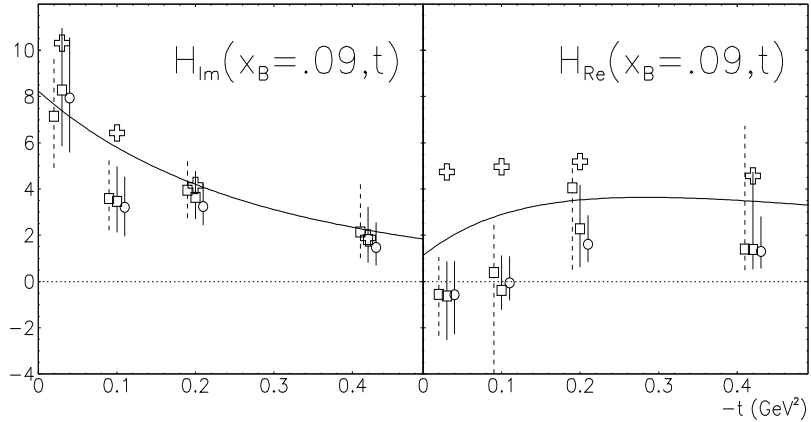}
\vspace{-5.5cm}
\caption{The $t$-dependence of the $H_\mathrm{Im}$ (left) and $H_\mathrm{Re}$
  (right) CFFs, extracted from our fits. The empty squares with the dashed 
  error bars are
  the results of Ref.~\cite{fithermes}, i.e.\ \underline{without} the
  new $A_\mathrm{UL}$ and $A_\mathrm{LL}$ moments in the fit (with 
  boundary values equal to 5 times the VGG reference values). The
  empty squares and circles with the solid line error bars are the results of the
  present work, i.e.\ with the additional moments of Ref.~\cite{hermesaul} in the fit 
  (though without the $A_\mathrm{\{Ux,I\}}^{\sin\phi}$ and $A_\mathrm{LL}$ moments,
  as discussed in the text). The squares (circles) correspond to
  boundary values 5 (3) times the VGG reference values. The empty crosses indicate
  the VGG prediction. The solid curves show the results of the
  model-based fit of Ref.~\cite{fitmuller}}
\label{fig:hermes4}
\end{figure}

Finally, we show in Fig.~\ref{fig:hermes4} how the $H_\mathrm{Re}$ and
$H_\mathrm{Im}$ CFFs have been affected by the introduction of the
$A_\mathrm{UL}$ and $A_\mathrm{LL}$ moments in the fit. The empty
squares with the dashed error bars are the results that we published
in Ref.~\cite{fithermes}, i.e.\ \underline{without} the new
$A_\mathrm{UL}$ and $A_\mathrm{LL}$ moments in the fit. The empty
squares with the solid line error bars are the results for the
$H_\mathrm{Re}$ and $H_\mathrm{Im}$ CFFs that we obtain in the present
work, i.e.\ with the addition of the $A_\mathrm{UL}$ in the fit (but without the
$A_\mathrm{\{Ux,I\}}^{\sin\phi}$ and $A_\mathrm{LL}$ moments as we discussed earlier).
The agreement between the ``dashed error bar" points and the ``solid line error bars" 
points is in general very good for both CFFs. There is almost no
difference for the $H_\mathrm{Im}$ CFF besides some reduction
in the error bar (most noticeable at the largest $<-t>$ value). The
reduction of the error bar is more significant and systematic for the
$H_\mathrm{Re}$ CFF. For this latter CFF, we also note a significant
change at $<-t>=0.201$ GeV$^2$, i.e. a lowering of the central value
(which is though still compatible with the ``dashed" error bar of our
previous study). In Fig.~\ref{fig:hermes4} we also show the VGG predictions
(empty crosses) and the result of the model-based fit of
Ref.~\cite{fitmuller} (solid curve) which was discussed in
Ref.~\cite{fithermes}. It is seen that both VGG and the model-based
fit of Ref.~\cite{fithermes} overestimate our fitted values for
$H_\mathrm{Re}$ while showing a good agreement for $H_\mathrm{Im}$.

To summarize this work, we have analyzed, in the leading twist/leading order 
handbag diagram and GPD framework, the new
longitudinally polarized target asymmetry data of the DVCS process,
recently released by the HERMES collaboration~\cite{hermesaul,davidthesis}. 
We have used a largely
model-independent fitter code, which has been introduced and used
succesfully in several previous
analyses~\cite{fitmick,fithermes,fittsa}, to fit these new data (in
addition to other DVCS observables previously published by the HERMES
collaboration).  We have met difficulties in fitting the
experimentally large $A_\mathrm{UL}^{\sin2\phi}$ moments which are
expected to be a higher twist effect. It thus cannot be described and
explained within our fitting framework. It should, however, be noted
that this large higher-twist effect is very local as it seems to stem
from only one out of four $x_B$ values. We have also met a difficulty
in fitting simultaneously the $A_\mathrm{UL}^{\sin\phi}$ and
$A_\mathrm{\{Ux,I\}}^{\sin\phi}$ moments at one particular $<-t>$ point
($<-t>=0.201$ GeV$^2$). This inconsistency, once again within our
leading twist DVCS formalism assumption, seems to be corroborated by
the VGG model which is in good agreement with the $A_\mathrm{UL}^{\sin\phi}$
moment but not with the $A_\mathrm{\{Ux,I\}}^{\sin\phi}$ moment. We note that,
in the comparison with the twenty-three independent moments measured
by HERMES, the VGG model gives a reasonable overall description of
the data.

These caveats being noted, our analysis has led for the first time
to some numerical constraints on the $\tilde{H}_\mathrm{Im}$ CFF at HERMES energies,
with well-defined and stable error bars and central values. Using a
previous work on CLAS data, we have presented the $\xi-$ dependence of
the $\tilde{H}_\mathrm{Im}$ CFF. It has then been observed that
$\tilde{H}_\mathrm{Im}$ exhibits the same peculiar $t-$ dependence at
both energies, i.e.\ a rather flat $t$-slope with, possibly,
a trend to decrease as $-t$ tends towards zero.
The $\xi-$ dependence is also very shallow with a slight decrease of
$\tilde{H}_\mathrm{Im}$ between the CLAS and HERMES kinematics. In addition, 
the results on the other $H_\mathrm{Re}$ and $H_\mathrm{Im}$ CFFs have 
also been improved with respect to our previous study~\cite{fithermes}.

We are very thankful to Drs D. Mahon, M. Murray, W.-D. Nowak, I. Lehmann and 
M. Vanderhaeghen for very useful discussions which have enriched this study.
This work was supported in part by the French "Nucleon GDR" no. 3034, 
the French Agence Nationale pour la Recherche Contract no. ANR-07-BLAN-0338 
and the EU FP7 Integrating Activity HadronPhysics2, in particular the Joint 
Research Activity HardEx.


\begin{thebibliography}{99}
\bibitem{muller}
D. M\"uller, D. Robaschik, B. Geyer, F.-M. Dittes, and J. Horejsi,
Fortschr. Phys. {\bf 42}, 101 (1994).
\bibitem{ji}
X. Ji, Phys. Rev. Lett. {\bf 78}, 610 (1997); 
Phys. Rev. D {\bf 55}, 7114 (1997).
\bibitem{rady}
A.V. Radyushkin, Phys. Lett. B {\bf 380} (1996) 417; 
Phys. Rev. D {\bf 56}, 5524 (1997).
\bibitem{collins}
J.C. Collins, L. Frankfurt and M. Strikman, 
Phys. Rev. D {\bf 56}, 2982 (1997).
\bibitem{goeke} K. Goeke, M. V. Polyakov and M. Vanderhaeghen,
Prog.\ Part.\ Nucl.\ Phys. {\bf 47}, 401 (2001).
\bibitem{revdiehl}
M. Diehl, Phys.  Rept. {\bf 388}, 41 (2003).
\bibitem{revrady}
A.V. Belitsky, A.V. Radyushkin, Phys. Rept. {\bf 418}, 1 (2005).
\bibitem{barbara}
S. Boffi and B. Pasquini, Riv. Nuovo Cim. {\bf 30}, 387  (2007).
\bibitem{myppnp} M. Guidal, Prog. Part. Nucl. Phys. {\bf 61}, 89 (2008).
\bibitem{fitmick} M. Guidal, Eur.\ Phys.\ J.\ A {\bf 37}, 319 (2008),
Erratum-ibid.A40:119,2009. 
\bibitem{fithermes} M. Guidal and H. Moutarde, Eur.\ Phys.\ J.\ A {\bf 42}, 
71 (2009).
\bibitem{fittsa} M. Guidal, Phys. Lett. B {\bf 689} (2010) 156.
\bibitem{kirch} A. Belitsky, D. Muller and A. Kirchner, Nucl. Phys. B {\bf 629}, 323 (2002).
\bibitem{vgg1} M. Vanderhaeghen, P.A.M. Guichon, M. Guidal, Phys. Rev. D {\bf 60}, 094017 (1999).
\bibitem{franck} C. Mu\~noz Camacho et al., Phys.\ Rev.\ Lett. {\bf 97}, 262002 (2006).
\bibitem{fx} F.-X. Girod et al., Phys.\ Rev.\ Lett. {\bf 100}, 162002 (2008).
\bibitem{chen} S. Chen et al., Phys.\ Rev.\ Lett. {\bf 97}, 072002 (2006).
\bibitem{ave} A. Airapetian et al., JHEP {\bf0806}, 066 (2008).
\bibitem{hermes} A. Airapetian et al., JHEP {\bf0911}, 083 (2009).
\bibitem{hermesaul} A. Airapetian et al., JHEP (in press) arXiv:1004.0177 [hep-ex].
\bibitem{davidthesis} D. Mahon, PhD thesis, University Glasgow (2010).
\bibitem{james} F. James, MINUIT, D507, CERN (1978).
\bibitem{gprv} M. Guidal, M. V. Polyakov, A. V. Radyushkin and 
M. Vanderhaeghen, Phys.\ Rev. D {\bf 72}, 054013 (2005).
\bibitem{Burkardt:2000}
M.~Burkardt, Phys.\ Rev.\ D {\bf 62}, 071503 (2000)
[Erratum-ibid.\ D {\bf 66}, 119903 (2002)] ;
Int.\ J.\ Mod.\ Phys.\ A {\bf 18}, 173 (2003).
\bibitem{Diehl:2002he}
M.~Diehl, Eur.\ Phys.\ J.\ C {\bf 25}, 223 (2002)
[Erratum-ibid.\ C {\bf 31}, 277 (2003)].
\bibitem{pire}
J.~P.~Ralston and B.~Pire, Phys.\ Rev.\ D {\bf 66}, 111501 (2002).
\bibitem{fitmuller} K. Kumericki and D. M\"uller, arXiv:0904.0458 [hep-ph].
\end{thebibliography}
\end{document}